# Persistence of an Energy Scale in Over-Doped High-$T_c$ Superconductors

T. Cuk*, A.D. Gromko[%], Zhe Sun[%], Z.-X. Shen*, D.S. Dessau[%]

*Dept of Applied Physics, Stanford University, Stanford, CA 94305*

*%Dept of Physics, University of Colorado, Boulder, CO 80309*

Citing the disappearance of a sharp peak in the electron self-energy, extracted from optics and angle-resolved photoemission spectroscopy (ARPES) experiments in deeply over-doped copper-oxide superconductors with $T_c$ of 55-60K, Hwang *et al.* claim that sharp modes, be they phononic or magnetic in origin, are not important for superconductivity in the cuprates[1]. If true, this would have been important progress. We show, however, their conclusions are unfounded because of the insensitivity of the optics experiment, and a misrepresentation of existing ARPES data.

Contrary to the claim of a *null* result, existing ARPES data show that, even in the deeply over-doped regime ($T_c$~58K), a prominent 'kink' directly indicating a peak in the self-energy exists, as in Fig.1[2]. Panels (a) and (b) show data from the antinodal region in the normal and superconducting state, respectively. The superconducting state data reveal a clear dispersion kink near 40 meV, as highlighted by the black arrow. The strong presence of the mode signal in this comparably overdoped sample invalidates the central claim by Hwang *et al.*

The conclusion of Hwang *et al.* is unfounded for two reasons. First, optics measures a momentum average and therefore is not a very sensitive probe when the signal is strongly momentum dependent. Figure 1d shows that the kink strength, or the peak height of the extracted self energy $\Sigma$, from an overdoped sample ($T_c$~71K) is hardly detectable near the node but is quite strong near the antinode, comparable to the maximum ARPES value quoted by Hwang et al. from deeply underdoped samples where the signal is the strongest. A signal easily detectable by ARPES when probing the appropriate momentum space region, as in Fig.1b, is missed in the optics measurement. Thus, Hwang *et al.* have failed to ensure that their experimental sensitivity is higher than the expected signal, a basic requisite to claim a negative result. Second, Hwang *et al.* selectively reference ARPES data taken at the node,[3] where the "kink" effect is the weakest as in Fig.1d[2,4,5], to confirm a *null* result; no mention is made of a clear *positive* ARPES signal at the antinode[2], which otherwise would have ruled out their conclusion. This omission is an oversight even within the context of their interpretation of the mode being the $(\pi,\pi)$ spin resonance at 41 meV, as a strong anisotropy of coupling to this mode is expected[6].

The presence and the dramatic enhancement of the energy scale (kink) below $T_c$ challenges the argument that it is not important to the pairing mechanism. The fact that the prominence of this energy scale decreases with doping attests to its importance to the pairing mechanism, as the superconducting gap, which represents the pairing strength, also decreases with doping and rapidly so in the over-doped regime. The possible candidates for the mode preferentially coupled to the anti-nodal states are the spin resonance[2-5] and the $B_{1g}$ phonon[7,8]. The relative merits of these two competing interpretations is not the subject of this comment, but the mode's presence and importance is unambiguous and invalidates the central claim by Hwang *et al*.

## Note Added in Reply to Valla *et. al.*[9]:

Recently, Valla, Timusk *et. al.*[9] have posted a reply to this comment[10] in defence of the conclusions made by Hwang, Timusk *et. al.*[1]. Instead of addressing the central point of the comment, the authors attempted to avoid the issue by arguing that the doping level of our sample is actually lower than reported, implying that our data is not relevant to the overdoped regime. However, Valla *et al.* have misrepresented our data, and using their same methods we show that there is no inconsistency among the data from all groups and our sample surface was clearly of the appropriate doping level.

In the reply, Valla *et. al.* defines the gap as the EDC peak position at the point in momentum space $(\pi, k_F)$. Yet, Fig. 1 of the reply reports the gap of our 58K sample to be 18meV, which is the energy of the EDC peak at $(\pi, 0)$ rather than $(\pi, k_F)$, as noted in Fig. 7 of Gromko *et. al.*[2]. The relevant data is presented as Fig. 1b and in Fig. 2a below. As seen from Fig. 2a), the actual energy at $(\pi, k_F)$ is 14 meV, which falls within the expected range of values for a sample of $T_c = 58K$.

Valla *et al*. did not specify how they extracted the vHS energy from the data. However, this is typically considered to be the energy of the broad "hump" of the normal-state EDC measured at $(\pi, 0)$. The data was shown by Gromko *et. al.*[2] and is reproduced as Figure 2b below. As shown in Fig. 2c, a two component fit (one for the "peak" and the other for the "hump") to the identical normal state data returns the value of 100 +/- 10 meV, instead of the reported value of 120 meV by Valla *et. al.* A value of 100 meV is consistent with what one could determine by inspecting the data of Fig. 2b by eye.

Therefore, we find no inconsistency between the doping level of our samples and a $T_c$ of 58K as claimed (see the revised Valla *et. al.* figure reproduced as Fig. 2d below).

Furthermore, the above two energy scales are from the identical spectrum which shows the kink. Using the criteria defined by Valla *et. al.*[9] and Hwang *et. al.*[1], it is therefore unambiguous that kinks in this heavily over-doped range exist.

The last sentence of the rebuttal states that "a similar conclusion" (lack of a kink for doping levels in excess of 0.23-0.24) has been made in the recent ARPES study of Kim *et. al.*[11]. Here we point out that the highest doping level from Kim *et. al*. was closer to 0.21 from where they did observe a kink (stated as a factor of 2 or more change in $\lambda$ between the normal and superconducting states). Extrapolating to the 0.23 doping range to claim a null effect is less convincing than real data (Gromko *et. al.*) from the doping level of interest which shows a clear positive effect.

The central criticism we report of Hwang *et. al.*'s experiment remains unaddressed: a) Valla *et al.* fail to demonstrate sufficient sensitivity to detect the signal, a fundamental requirement of any null claim; b) the momentum averaged optics technique is inadequate to address these issues when the momentum dependence is strong, as in this case.

1. Hwang, J., Timusk, T., and Gu, G.D., *Nature* **714**, 412-717 (2004).

2. Gromko, A.D. et al. *Phys. Rev. B* **68**, 174520 (2003).

3. Johnson, P.D. et al., *Phys. Rev. Lett.* **87**, 177007 (2001).

4. Kaminski, A. et al., *Phys. Rev. Lett.* **86**, 1070 (2001).

5. Sato, T. et al., *Phys. Rev. Lett.* **91**, 157003 (2003).

6. Abanov, Ar. et al., *Phys. Rev. Lett.* **89**, 177002 (2002).

7. Cuk, T. et al., submitted to *Phys. Rev. Lett.*, cond-mat/0403521 (2004).

8. Devereaux, T. et al., submitted to *Phys. Rev. Lett.*

9. Valla, T. et. al., xxx.lanl.gov, cond-mat/0405203.

10. Cuk, T. et. al., xxx.lanl.gov, cond-mat/0403743.

11. Kim, T. K. et. al., *Phys. Rev. Lett.* **91**, 167002/1-4 (2003).

12. Campuzano, J.C. et. al., *Phys. Rev. Lett.* **83**, 3709 (1999).

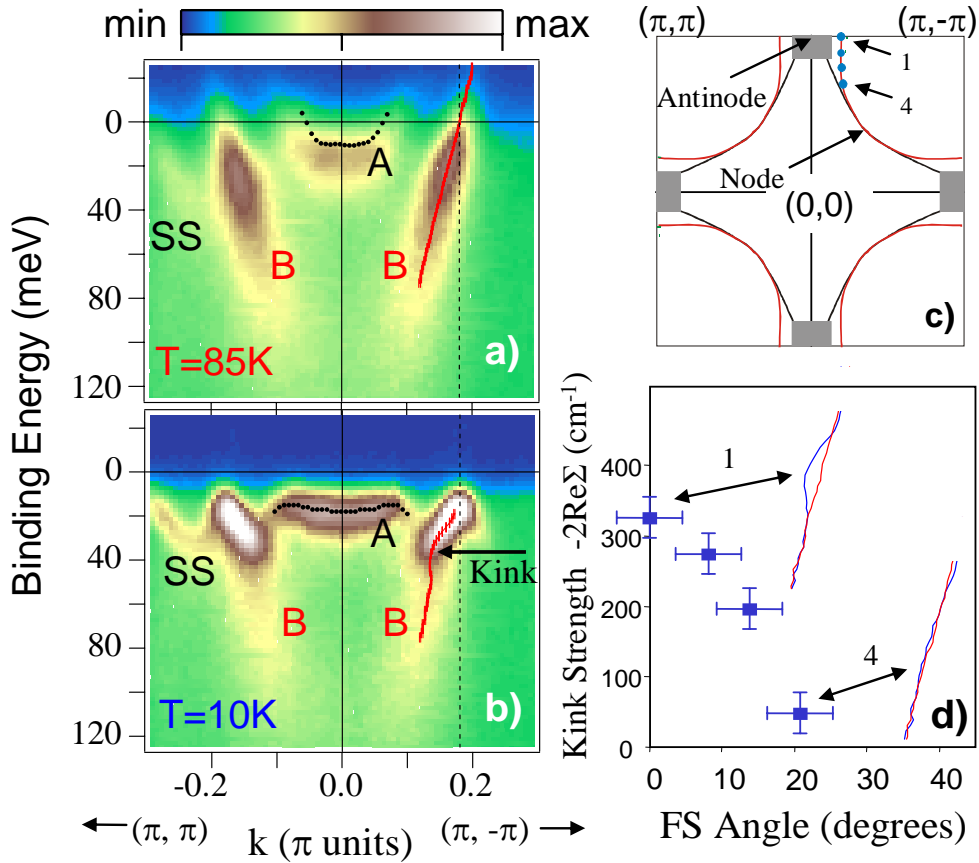

**Figure 1**. ARPES data showing a kink in a heavily overdoped Bi2212 sample, after ref. 2. (a) Normal state data from an OD $T_c$=58K samples near the anti-nodal region (see panel c). (b) Superconducting state data from the same sample showing the emergence of a dispersion kink in the bilayer split B band (see black arrow). (d) Momentum dependence of the strength of the temperature dependent kink (the real part of the self energy $\Sigma$, taking the normal state curve as a reference) from an over-doped $T_c$=71K sample, with locations 1 through 4 indicated on the Brillouin zone of panel (c). The normal (red) and superconducting (blue) dispersion curves for locations 1 and 4 are shown as well. The ARPES spectra discussed in ref. 1 were taken at 45 degrees (e.g. the node).

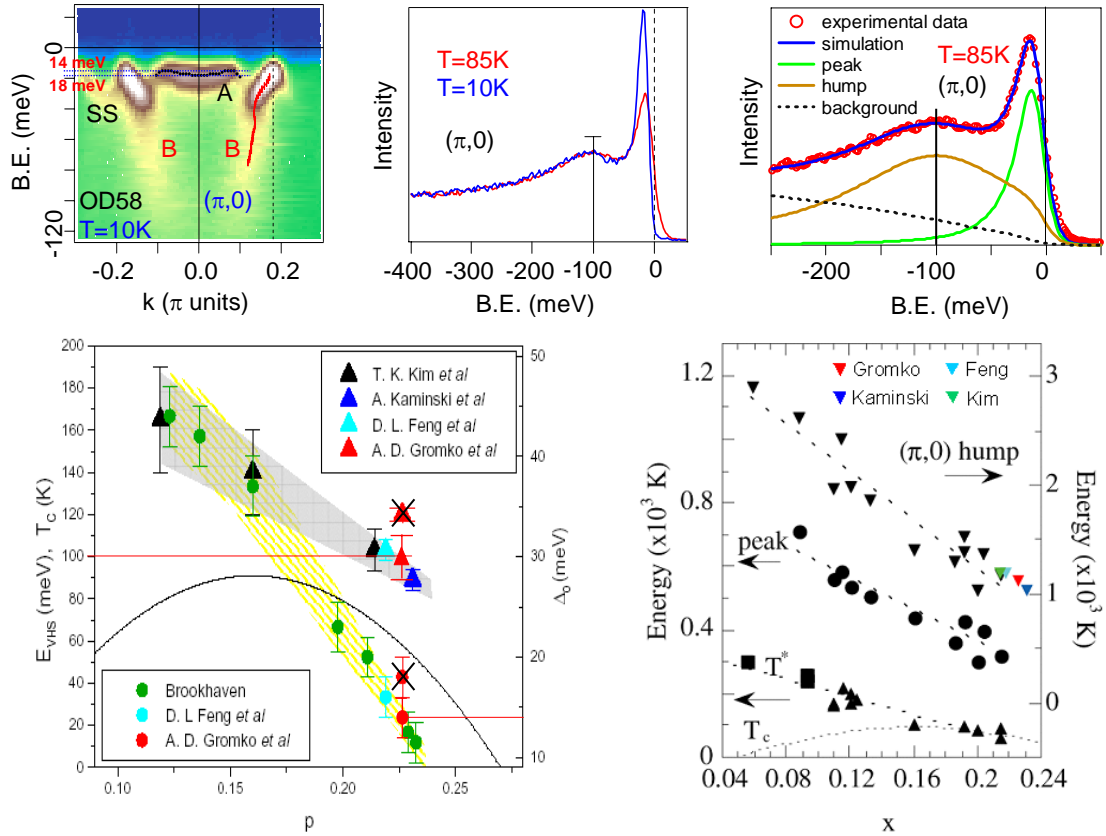

**Figure 2**. (a) Reproduction of Figure 2b from Gromko *et. al.* showing the dispersion of the EDC peak (black dots). While the peak at $(\pi,0)$ is 18meV, it is 14meV at $(\pi,k_F)$ which is the appropriate scale for comparison. (b) $(\pi,0)$ EDC from Gromko *et al.* showing the vHS energy of ~ 100meV. (c) Two-component fit to the same normal state data, returning a value of 100meV. (d) Reproduction of Figure 1 from Valla *et. al.* with the correct energy scales assigned to the Gromko *et. al.* data. (e) Summary of vHS energies from all groups shown on a reproduction from Campuzano *et. al.*[12]